\documentclass[11pt,twoside]{article}
\usepackage{asp2004}
\usepackage{psfig}
\usepackage{epsf}
\usepackage{graphicx}
\usepackage{bm}
\usepackage{lscape}

\newcommand{\Fig}[1]{Fig.~\ref{#1}}
\newcommand{\bra}[1]{\langle #1\rangle}
\newcommand{\meanBB}{\overline{\bm{B}}}
\newcommand{\BB}{{\bm{B}}}
\newcommand{\G}{\,{\rm G}}
\newcommand{\kG}{\,{\rm kG}}
\newcommand{\Mm}{\,{\rm Mm}}

\newcommand{\yapj}[3]{ #1, {ApJ,} {#2}, #3}
\newcommand{\yapjl}[3]{ #1, {ApJ,} {#2}, #3}
\newcommand{\yan}[3]{ #1, {AN,} {#2}, #3}
\newcommand{\yana}[3]{ #1, {A\&A,} {#2}, #3}
\newcommand{\ygafd}[3]{ #1, {Geophys. Astrophys. Fluid Dyn.,} {#2}, #3}
\newcommand{\ysov}[3]{ #1, {Sov. Astron.,} {#2}, #3}
\newcommand{\yprl}[3]{ #1, {PRL,} {#2}, #3}

\newcommand{\ynat}[3]{ #1, {Nat,} {#2}, #3}

\newcommand{\ysph}[3]{ #1, {Sol. Phys.,} {#2}, #3}
\newcommand{\ypre}[3]{ #1, {PRE,} {#2}, #3}
\newcommand{\yjour}[4]{ #1, {#2}, {#3}, #4}

\def\edcomment#1{\iffalse\marginpar{\raggedright\sl#1\/}\else\relax\fi}
\reversemarginpar

\begin{document}
\title{Location of the solar dynamo and near-surface shear}
 \author{A. Brandenburg}
\affil{Nordita, Blegdamsvej 17, D-2100 Copenhagen \O, Denmark}

\begin{abstract}
The location of the solar dynamo is discussed in the context of
new insights into the theory of nonlinear turbulent dynamos.
It is argued that, from a dynamo-theoretic point of view, the
bottom of the convection zone is not a likely location
and that the solar dynamo may be distributed over the convection zone.
The near surface shear layer produces not only east-west field alignment,
but it also helps the dynamo disposing of
its excess small scale magnetic helicity.
\end{abstract}

\vspace{-0.5cm}
\section{Introduction}

\begin{figure}[t!]\centering
\includegraphics[width=0.70\textwidth]{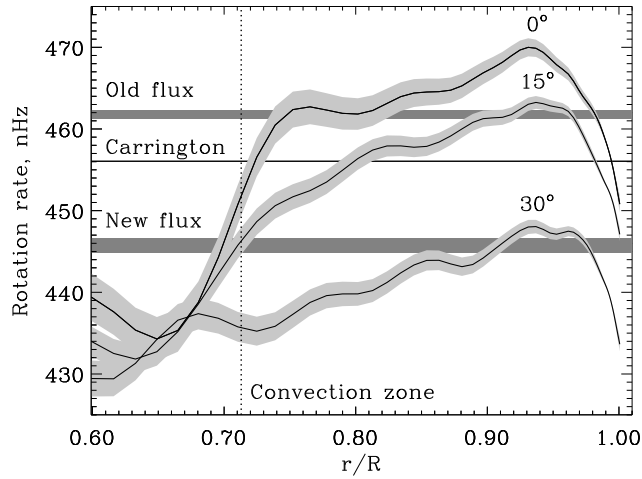}\caption{
Radial profiles of the internal solar rotation rate, as inferred from
helioseismology.
The rotation rate of active zones at the beginning of the cycle
(at $\approx30^\circ$ latitude) and near the end (at $\approx4^\circ$)
is indicated by horizontal bars, which intersect the profiles of
rotation rate at $r/R_\odot\approx0.97$.
Courtesy of Benevolenskaya et al.\ (1999).
}\label{bene99}\end{figure}

It is commonly taken for granted that the solar dynamo has to work
at the bottom of the convection zone, or that at least the toroidal
field is generated or stored down there (Spiegel \& Weiss 1980,
Golub et al.\ 1981, Galloway \& Weiss 1981, Choudhuri 1990).
This expectation results mostly from the fact that only at the bottom
of the convection zone the dynamical time scales associated with
convection and magnetic buoyancy are long enough to be comparable with
the rotational period.
There is also the notion that the magnetic field needs to be `stored'
over a significant fraction of the solar cycle period and that this
is only conceivable at or below the base of the convection zone.
There are several other aspects in favor of placing the dynamo at the
bottom of the convection zone.
One is the large extent of the active regions (up to $100\Mm$) that
is only compatible with length scales typical of the deep convection
zone (Galloway \& Weiss 1981).
Another argument is that it is at the bottom of the convection zone
that we have a strong radial shear layer where
$r\partial\Omega/\partial r\neq0$.
However, there is of course also latitudinal differential rotation
($\partial\Omega/\partial\theta\neq0$) that is actually stronger, and
there is still extremely strong radial shear just beneath the surface
in the uppermost $30\Mm$ of the Sun (see \Fig{bene99}).
So, we see that the shear argument is problematic.
In addition, at the bottom of the convection zone the sign of the radial
shear is such that standard dynamo theory would predict an
equatorward migration only when the $\alpha$ effect is negative.
Very near the bottom of the convection zone the $\alpha$ effect is
indeed predicted to have the opposite sign according to the
standard formalism (Krivodubskii 1984).
However, there is a whole host of other problems.
First of all, the radial shear seen at the bottom of the convection
zone is strongest at the poles and this is also where $\alpha$
is strongest.
So, in spite of spherical geometry factors the magnetic activity
predicted by overshoot layer dynamos is far too strong at the
poles and needs to be artificially suppressed if this approach
is to be viable (R\"udiger \& Brandenburg 1995, Markiel \& Thomas 1999).
Secondly, such overshoot layer dynamos (also sometimes called tachocline
dynamos) have the well-known problem of producing too many toroidal
field belts in the meridional plane (Moss et al.\ 1990).
Furthermore, the negative radial angular velocity gradient in the
bulk of the convection zone and especially at the bottom tends to
produce the wrong migration direction of the magnetic activity,
i.e.\ poleward rather than equatorward (Parker 1987).
Although this problem could be fixed by invoking a strong negative
value of $\alpha$ at the bottom of the convection zone,
there remains always the problem with the phase relation between
radial and azimuthal fields, i.e.\ $B_rB_\phi$ is observed to be
negative, but it would be positive with positive radial shear
(Yoshimura 1976, Stix 1976).

Even if one ignored all these problems, there are still a number of
difficulties associated with the idea of having a dynamo operating at
the bottom of the convection zone.
Firstly, in order for the flux tubes
to be correctly oriented after their ascent,
the field strength of the flux tube has to be very
strong ($\sim100\kG$) to resist extraordinarily strong distortions and tilt.
However, it is hard to imagine that the field strength exceeds the
equipartition value ($\sim1\kG$) by a factor of a hundred, and this has
not yet been demonstrated.
Secondly, it is hard to imagine that the flux tubes would not disrupt
by expanding too much before forming a neat sunspot pair.

These are problems and difficulties that we have been living with
for quite a few years when constructing overshoot layer dynamos.
However, there is also the possibility of placing the dynamo right in
the middle of the convection zone.
This idea may appear rather uncomfortable at first, but to people
working in dynamo theory it is a rather natural and appealing scenario.
The basic picture is one where dynamo action occurs in the bulk of the
convection zone, affected obviously by the near-surface shear layer.
Downward pumping will also operate, so as to prevent the magnetic
field from floating upwards on too short a time scale
(Nordlund et al.\ 1992, Tobias et al.\ 1998).
However, in this scenario
the field that we observe as sunspots at the surface is likely to
come from the near surface layers, where sunspots may form as a result of
convective collapse of magnetic fibrils (Zwaan 1978, Spruit \& Zweibel 1979),
possibly facilitated by negative turbulent magnetic pressure effects
(Kleeorin et al.\ 1996) or by an instability (Kitchatinov \& Mazur 2000)
causing the vertical flux to concentrate into a tube.
The anticipated averaged field strength in the convection zone
would be about $300\G$, i.e.\ about $10\%$ of the equipartition value.
This field may then get amplified locally near the surface.
In that sense, sunspots are not be deeply rooted, but
rather a shallow phenomenon rooted at a depth of $20$--$30\Mm$.

\section{Distributed dynamos with shear}

In this section we discuss some of the properties of turbulent dynamos
that are strongly affected by shear.
Some of the recent findings have been presented in an earlier paper
(Brandenburg 2005; hereafter B05).
Here we will only review the main aspects.
Before going into details, it is important to put this research into
perspective.
Let us distinguish three different aspects of dynamos.
There are first of all the mean field dynamos, which is based on a
theory for the averaged magnetic field, whose evolution is dominated by
parameters such as $\alpha$ effect and turbulent diffusivity.
Without any independent confirmation of the existence and magnitude
of these coefficients, the predictive power of this approach is limited.
We shall not be concerned with this approach in any details, except for
the comparison with other approaches.
Then there are dynamos where turbulence is not parameterized, but it
is explicitly being solved for using computer simulations at the
highest possible resolution.
Two types of these dynamos can be distinguished: small scale and large
scale dynamos.
Both are turbulent and both have small scale magnetic fields, but the
large scale dynamos also show large scale spatial coherence, and
in some cases even long term temporal coherence such as magnetic cycles.
The latter type of dynamo is clearly relevant to the Sun, while the
former one may be dominant in many simulations.

A general remark is here in order.
In many simulations the magnetic Prandtl number (ratio of
viscosity to magnetic diffusivity) is of order unity, while in the
Sun it is $<10^{-4}$.
Only in recent years the possible significance has been clarified.
It turns out that for progressively smaller magnetic Prandtl numbers
the threshold for dynamo action moves to larger magnetic
Reynolds numbers.
In the paper by Haugen et al.\ (2004) it was found that, for a
limited parameter range, the critical magnetic Reynolds numbers
scales with the magnetic Prandtl number to the $-1/2$ power.
However, in reality this dependence may actually not be a power law and
there are now suggestions that the slope may become steeper toward smaller
values of the magnetic Prandtl number, and that small scale dynamos may
even become completely impossible below a certain critical value
(Schekochihin et al.\ 2005).
At the same time, the large scale dynamo is largely independent of
magnetic Prandtl number, as will be discussed next.

A prime example of a large scale turbulent dynamo is in the presence
of helicity.
In this case the magnetic Reynolds number, defined here as
$R_{\rm m}=u_{\rm rms}/(\eta k_{\rm f})$, where $u_{\rm rms}$ is the
rms velocity, $\eta$ the magnetic diffusivity, and $k_{\rm f}$ the
typical wavenumber where the kinetic energy spectrum peaks.
Both for $P_{\rm m}=1$ and for $P_{\rm m}=0.1$ the critical value
of $R_{\rm m}$ for large scale dynamo action is around unity
[see Table~1 of Brandenburg (2001), but there the magnetic Reynolds
numbers need to be divided by $2\pi$ to comply with the definition above].
If the absence of small scale dynamo action for small magnetic Prandtl numbers
is confirmed, this might suggest that in bodies such as the Sun, only
large scale dynamo action is possible, but not small scale dynamo action.
Alternatively, the naive extrapolation to solar parameters may be
invalid, so it is possible that for sufficiently small values of the
magnetic Prandtl number the critical value of $R_{\rm m}$ for small scale
dynamo action levels off at a constant value of perhaps several hundred
(Boldyrev \& Cattaneo 2004).
However, such high values are currently still out of reach for direct
simulations.

\begin{figure}[t!]\centering
\includegraphics[width=0.80\textwidth]{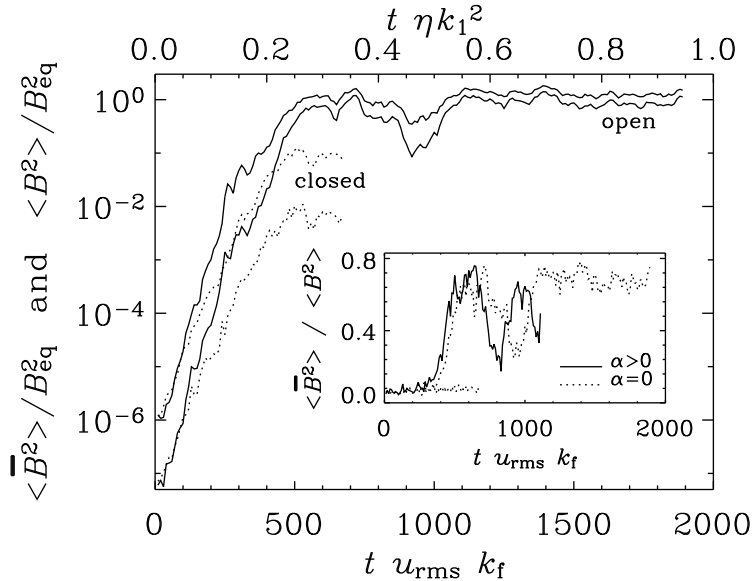}\caption{
Evolution of the energies of the total field $\bra{\BB^2}$ and of
the mean field $\bra{\meanBB^2}$, in units of $B_{\rm eq}^2$,
for runs with non-helical forcing
and open or closed boundaries; see the solid and dotted lines, respectively.
The inset shows a comparison of the ratio $\bra{\meanBB^2}/\bra{\BB^2}$
for nonhelical ($\alpha=0$) and helical ($\alpha>0$) runs.
For the nonhelical case the run with closed boundaries is also shown.
Saturation of the large scale field occurs on a
dynamical time scale; the resistive time scale is given on the
upper abscissa. [Adapted from B05.]
}\label{pmean_comp}\end{figure}

The significance of these considerations is that, when trying to find
solar-like dynamo action on the computer, it is not enough to find that the
magnetic field is growing.
Instead, the field should also be of large scale.
This may not be the case, even for the currently best resolved dynamos
in full global spherical shell geometry (Brun et al.\ 2004).
Large scale and small scale dynamo action are in this sense quite
different phenomena with different excitation conditions.

\begin{figure}[t!]\centering
\includegraphics[width=0.62\textwidth]{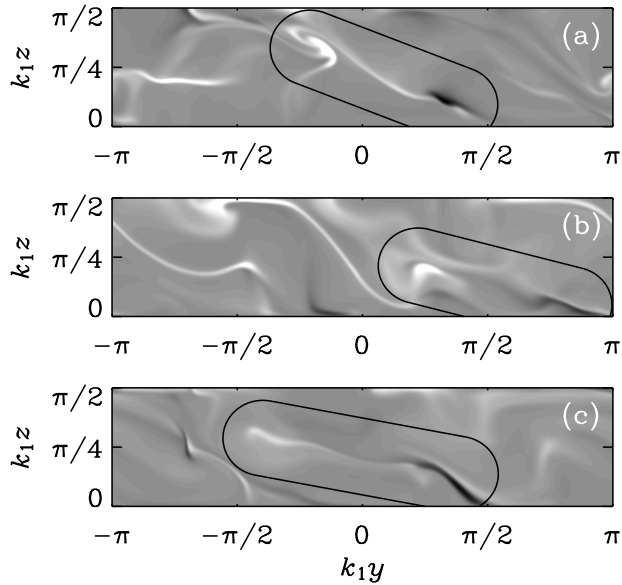}\caption{
Magnetograms of the radial field at the outer surface
on the northern hemisphere at different times.
Light shades correspond to field vectors pointing out of the domain,
and dark shades correspond to vectors pointing into the domain.
The elongated rings highlight the positions of bipolar regions.
Note the clockwise tilt relative to the $y$ (or toroidal) direction
of systematically aligned bipolar regions. [Adapted from B05.]
}\label{pmagnetogram}\end{figure}

We have already mentioned that large scale dynamo action is possible for
helical turbulence.
The qualitative picture is well understood in the framework of mean field
dynamo theory, and even its nonlinear saturation behavior is
well reproduced in the absence of boundaries (Field \& Blackman 2002).
Thus, in sufficiently simple situations, such as these, mean field theory
does actually begin to have predictive power.
Even in the presence of shear, the theory with dynamical
quenching formalism predicts a nonlinear behavior that is compatible with
simulations (Blackman \& Brandenburg 2002, Subramanian 2002).

Shear leads to two important effects.
The first one was long known: the amplification of a toroidal field
from a poloidal one.
The second is far less obvious and has only recently been discussed: the
transport of magnetic and current helicity along lines of constant angular
velocity (Vishniac \& Cho 2001, Subramanian \& Brandenburg 2004).
Qualitatively, any large scale dynamo, even if the turbulence is not
driven helically, does imply the production of small scale magnetic
and current helicities, which tends to `suffocate' the large scale
dynamo process (Brandenburg et al.\ 2002).
In order to prevent this from happening, it is important to expel
small scale magnetic and current helicity, e.g.\ via transport
along linear of constant angular velocity, which is why this shear
is so important.
That this actually makes a tremendous difference becomes clear from
\Fig{pmean_comp}, where we show the growth of magnetic energy contained
in the total field, $\bra{\BB^2}$, and the mean field, $\bra{\meanBB^2}$.
Here, overbars denote averages over a meridional plane and angular
brackets denote volume averages.
All models have the same amount of shear, some models have helical
forcing of the turbulence while others have non-helical forcing;
this does not make a big difference as far as the generation of
magnetic energy is concerned (cf.\ solid and dashed lines in the inset
of \Fig{pmean_comp}).
Furthermore, there is one model that has closed boundaries -- preventing
the generated magnetic and current helicities to escape (dashed line in
main part of plot).
The effect is dramatic!
Both large scale and small scale fields saturate at a level that is
well below the equipartition value -- in stark contrast to the case with
open boundaries.

Finally we discuss the topology of the
generated magnetic field, as viewed at the outer surface of the domain;
see \Fig{pmagnetogram}.
The magnetic field appears to be tube-like at the outer surface, even though
meridional cross-sections of the field show a rather smooth field distribution
(see Fig.~4 of B05).
This suggests that the localized appearance the field is
primarily produced close to the boundary.
Furthermore, the field appears in the form of bipolar regions
with a systematic tilt angle.
Here the tilt is produced by latitudinal shear, which causes
all points closer to the equator to drift faster than those further
away (see B05 for details).

\section{Conclusion}

The main point of this discussion is to stress that the solar dynamo
may well work in the bulk of the convection zone.
The near surface shear may not only be responsible for east-west
alignment and toroidal field production, but it may also play a role
in disposing of small scale magnetic and current helicities from the dynamo,
e.g.\ via coronal mass ejections (Blackman \& Brandenburg 2003).

%\acknowledgements %%% Text of acknowledgements runs on after this command.

%%% THE BIBLIOGRAPHY
%%%
%%% CONSULT SECTION 3 OF "INSTRUCTIONS FOR AUTHORS" FOR HOW TO USE NATBIB.
%%% AUTHORS ARE ENCOURAGED TO USE EITHER THE "THEBIBLIOGRAPY" ENVIRONMENT
%%% BY UNCOMMENTING (DELETING THE "%" SYMBOL) THE COMMANDS BELOW, OR BY
%%% USING THE BIBTEX ENVIRONMENT. TO FIND OUT WHICH IS APPLICABLE TO YOUR
%%% CONTRIBUTION, CONSULT THE VOLUME EDITORS FOR YOUR PROCEEDINGS.
%%%

\end{document}